# How will advanced AI systems impact democracy?

Christopher Summerfield[1]*, Lisa Argyle[2], Michiel Bakker[3], Teddy Collins[4], Esin Durmus[5], Tyna Eloundou[6], Iason Gabriel[3], Deep Ganguli[5], Kobi Hackenburg[7], Gillian Hadfield[8], Luke Hewitt[9], Saffron Huang[4], Helene Landemore[10], Nahema Marchal[3], Aviv Ovadya[11], Ariel Procaccia[12], Mathias Risse[13], Bruce Schneier[13], Elizabeth Seger[14], Divya Siddarth[4], Henrik Skaug Sætra[15], MH Tessler[3], Matthew Botvinick[16].

[1] Department of Experimental Psychology, University of Oxford, Anna Watts Building, Radcliffe Observatory Quarter, Woodstock Road, Oxford, OX2 6GG
[2] Department of Political Science, Brigham Young University, 745 KMBL, Provo UT 84604, USA
[3] Independent contributor
[4] Collective Intelligence Project, 3411 Silverside Road, Tatnall Building 104, Wilmington, DE, USA
[5] Anthropic, 731 Sansome Street, 5th Floor, San Francisco CA 94104, USA
[6] OpenAI, 3180 18th St., San Francisco, CA 94110, USA
[7] Oxford Internet Institute, University of Oxford, 1 St Giles, Oxford, OX1 3JS, UK
[8] Faculty of Law, University of Toronto, Jackman Law Building, 78 Queen's Park, Toronto, Ontario M5S 2C5, Canada
[9] Stanford Center on Philanthropy and Civil Society, 559 Nathan Abbott Way, Stanford, CA 94305, USA
[10] Department of Political Science, Yale University, 115 Prospect Street, Yale, NH, USA
[11] AI & Democracy Foundation, 440 N Barranca Ave #8874 Covina, CA 91723, USA
[12] School of Engineering and Applied Sciences, Harvard University, 150 Western Avenue, Allston, MA 02134, USA
[13] Harvard Kennedy School, Harvard University, 79 John F. Kennedy St, Cambridge, MA 02138, USA
[14] Demos, 15 Whitehall, London, SW1A 2DD, UK
[15] University of Oslo, Department of Informatics, 0373 Oslo, Norway
[16] Yale Law School, 127 Wall St, New Haven, CT 06511, USA

* Corresponding author

## Abstract

Advanced AI systems capable of generating humanlike text and multimodal content are now widely available. In this paper, we discuss the impacts that generative artificial intelligence may have on democratic processes. We consider the consequences of AI for citizens' ability to make informed choices about political representatives and issues (epistemic impacts). We ask how AI might be used to destabilise or support democratic mechanisms like elections (material impacts). Finally, we discuss whether AI will strengthen or weaken democratic principles (foundational impacts). It is widely acknowledged that new AI systems could pose significant challenges for democracy. However, it has also been argued that generative AI offers new opportunities to educate and learn from citizens, strengthen public discourse, help people find common ground, and to reimagine how democracies might work better.

# Introduction

In 2024, half the world – including India, the US and several other of the world's richest and most populous nations – go to the polls. In the short intervening time since these voters last cast their ballot, there has been a step change in the development of advanced AI systems. Large generative models can now produce text, images, audio, and audio-visual outputs that closely resemble those produced by humans. In November 2022, OpenAI launched a publicly available large language model (LLM) on its ChatGPT website, which is now the 20$^{th}$ most visited internet page globally, flanked by rival systems from Google (originally Bard; now Gemini), Anthropic (Claude) and similar open-source versions. Over recent months, the impact that these powerful, publicly available AI systems may have on the political process has been widely debated in the media, often with a focus on the potential of AI to disrupt or corrode democracy. Here, we situate this discussion in a growing academic literature across both AI research and social science[1–7].

Democracy is a system of government in which the people, rather than monarchs or oligarchs, hold political power. In modern liberal democracies this power is mainly exercised through the act of voting for political representatives, although some democracies also empower mass decision-making through referenda. Different conceptions of democracy emphasise the struggle among leaders to gain the popular vote[8], how interest groups and organisations seek to influence the political process[9] or the processes by which political decisions are made (e.g., voting, negotiation, or deliberation)[10]. Language plays an indispensable role in each of these conceptions of democracy. It allows information about candidates and policies to be shared with to voters, lawmakers to create legislation, and citizens and representatives to collectively discuss, deliberate and decide on which course of action to pursue. In liberal democracies, elected leaders use oratory to explain and justify their decisions and actions to the larger public, interest groups use persuasive messaging to lobby for their preferred policies, and the general public engages in debate, either informally or through organised events such as town halls and citizens' assemblies[11,12]. Given the primacy of linguistic exchange in the political process, the arrival of conversational machines – such as ChatGPT, which is already generating more than 100 billion words per day[13] – has the potential for far-reaching impact on democracy worldwide.

We propose that AI creates three classes of potential challenge for democracy, but argue that each is tempered by corresponding potential opportunities. First, we consider *epistemic* impacts – those that impact citizens' ability to make informed choices about both representatives and policies. There is widespread concern that LLMs will spread misinformation at scale, or be used to craft highly persuasive political messages that undermine voters' ability to make autonomous decisions in their own interest. However, there is also considerable scope for AI to improve the epistemic health of our democracies, by providing voters with accurate and balanced information about political events, policies or leaders, by automating fact-checking, or helping people deliberate and find common ground over principles and issues.

Secondly, we turn to *material* impacts. AI could be misused to attack the infrastructure that supports democratic procedures, for example by overwhelming electoral processes or unfairly disenfranchising voters. However, it can also be deployed to improve the efficacy of governance processes, by helping policymakers make better use of data, or providing citizens with accurate information about their rights.

Finally, we discuss *foundational* impacts, by which AI may weaken or strengthen the very principles on which democracy is based, or affect its opportunity to flourish worldwide. Foundational impacts, whilst potentially mediated by epistemic or material impacts, have more diffuse, systemic, and long-lasting effect. Threats to the foundations of democracy could arise if AI serves to reduce human accountability[14], marginalise minorities[15], increase untoward surveillance or other forms of state-backed oppression[16], or shift the balance of power and wealth in ways that undermine the ideals of freedom, equality and inclusivity on which the democratic process is based[17]. However, new AI technologies could potentially be used to dramatically improve the democratic experience for citizens, for example by assimilating diverse perspectives at scale in natural language. This could open new doors for democratic participation, and perhaps help alleviate the current sentiment among many voters that they are ignored or 'left behind'[18].

## Epistemic impacts

Even before powerful LLMs became available, algorithms were responsible for shaping the flow of information and misinformation on digital platforms[19]. Algorithm design has often been blamed for the erosion of public discourse on social media, and for growing polarisation and partisanship in political debate[20]. However, the advent of LLMs presages new challenges and opportunities for global epistemic health. We consider how democracies may be weakened by political bias in AI systems, automated persuasion, polarisation from personalised content, or the scaling misinformation; but also how they may be strengthened by AI systems that allow fact-checking, increased mutual intelligibility, deliberative upskilling, and automated tooling for political participants to find common ground.

**Political bias**

Publicly available LLMs already have wide user bases, thought to collectively exceed 100 million monthly users. If citizens are using LLMs such as ChatGPT, Gemini or Claude to obtain information about current affairs, political controversies and electoral choices, then even weak biases in their outputs could significantly impact the distribution of political beliefs in this population. Several studies have attempted to quantify the degree of LLM political bias, typically by administering multiple choice survey questions (such as the Political Compass test[1]) to LLMs, and measuring the relative output probability associated with each candidate answer (e.g. option A vs. option B). These studies have shown that models are broadly calibrated to the distribution of political views in their training data, so that after pre-training on large datasets, LLMs reflect a wide spectrum of opinions, encompassing both more conservative and more liberal perspectives. However, when the multiple choice approach is applied to models that have undergone certain forms of fine-tuning, designed to minimise toxic or illegal outputs in the models[21], models have been reported to prefer options that tend in a more libertarian (e.g., favouring deregulation) and progressive (e.g., supporting civil rights) direction[22–24,25]. However, at the same time, LLMs are highly malleable, and when prompted to play the role of characters with different political opinions and worldviews, they are quick to adopt political opinions of both Republicans and Democrats[26,27]. Moreover, subsequent research has revealed that this stylised evaluation method, whereby the model is forced to choose a candidate response to an issue-based question, yields results that are unrepresentative of everyday user interactions with an LLM, because multiple choice items do not offer respondents the opportunity to voice balanced or equivocal replies. In fact, when responding freely to user queries in everyday settings, models like ChatGPT typically preface replies with reminders that they do not hold political opinions, and give scrupulously balanced answers to direct enquiries about the relative merits of political representatives or policies. They also remind the user about the limits of their knowledge, and refer them to sources on the internet for the most up-to-date information. In fact, under normal usage conditions, the models may be much less opinionated than previously argued[28]. Major public LLMs will often refuse to take sides on controversial political questions, a tendency which has led to complaints that models are respecting the censorship of oppressive regimes, or giving false equivalence to aggressor and victim on fraught question such as the war in Ukraine or the persecution of the Uyghurs[29]. Developers face difficult choices when choosing training protocols, being obliged to decide when models should take sides, and when it should acknowledge potentially conflicting perspectives. This has led some to seek public input to these questions[30,31].

---

[1] https://www.politicalcompass.org/test

**Persuasive messaging and dialogue**

During an election campaign, candidates, parties and interest groups attempt to shape voters' beliefs through advertisements, media engagement, public events, door-to-door canvassing, and other activities. Several recent studies, focussed on the US and UK electorate, have aimed to directly measure the impact of LLM-generated messaging on political attitudes through randomised controlled trials. We summarise their results in Figure 1.

First, studies have consistently found that LLMs are able to write messages that persuade on political issues. For example, messages crafted by GPT-3 increased support among a representative sample of US voters for a ban on smoking, or a tightening of gun control policy, by about 2-4% on average[32]. However, when comparing LLMs against human-written messages, research findings have been more mixed. In one study, messages generated by GPT-4 were significantly more persuasive than those written by experts such as political consultants[33], whereas another found that messages generated by Claude 3 Opus were no more persuasive than those written by laypeople[34]. These findings suggest that, at present, LLMs' greatest potential impact is to cheaply and rapidly produce persuasive content at a roughly human level[35], rather than to substantially improve upon the impact of campaigns' messaging itself. Given the pace of research in the field, however, this picture may well change with the release of new models, or the development of new prompting approaches.

Next, we consider another potentially significant capacity of LLMs: their use in political microtargeting. This is the practice of tailoring political messaging to a specific individual, based on such features as their demographic data or social media activity[36,37], which became notorious after Cambridge Analytica scraped data from 50 million Facebook users to target political ads during the 2016 US presidential election and UK Brexit referendum. On the one hand, microtargeting could bring benefits to the political process. By tailoring messages to voters' concerns, it may increase participation[38], or heighten interest in topics relevant to minority voters[39]. However, there is concern that LLMs could distort campaigning by mass-producing highly tailored messages with minimal human intervention. Studies have already shown that LLMs can infer political preferences from a small snapshot of user data, such as a single tweet[40], and LLM-generated messages are viewed as more persuasive when tailored to participants' personality traits[37,41]. However, as shown in Fig. 1, none of the three studies which directly measured the effect of targeted messaging on participant's attitudes showed a significant difference between the impact of targeted and untargeted LLM messages[42]. These findings align with existing work suggesting that microtargeted messages are rarely more effective than the single most persuasive message across the entire population[43], and suggest that at present, the use of LLMs for tailored political messaging may be less transformative than has been feared.

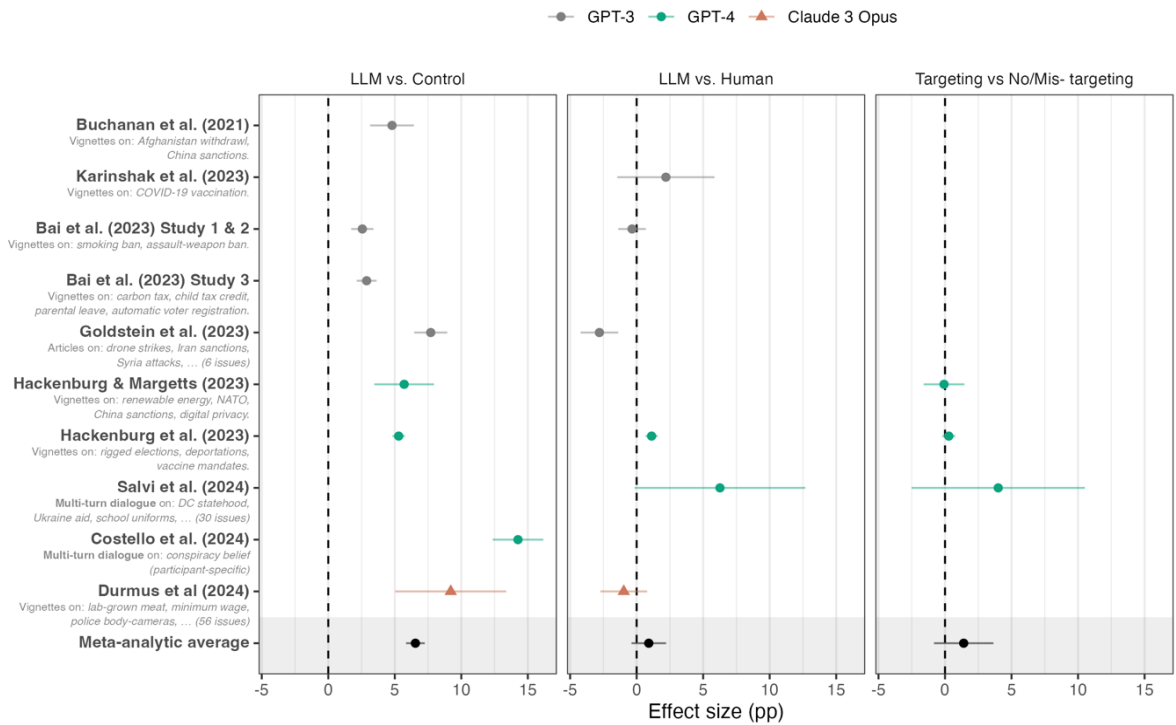

**Figure 1. RCT estimates of political persuasion with large language models**. This figure includes all known studies which randomised participants to LLM-generated political messages and measured post-treatment attitudes. For each study, we calculate the simple difference in mean outcomes by condition (with 95% CIs) in order to maximise consistency across studies, but note that this may differ from authors' original analyses. The studies vary in the model used (GPT-3, GPT-4, Claude 3 Opus), treatment format (vignettes, articles, chatbot conversations), reference conditions (experts, laypeople, etc.), as well as in the political issues considered. For descriptive purposes we include a meta-analytic average, but caution against over-interpretation given the substantial heterogeneity.

Interpersonal dialogue may be a more effective way to shift political views than static messages. For example, doorstep canvassing was able to durably change attitudes to transgender rights or immigration[44]. Given the ease of prompting LLMs to engage in persuasive, multi-turn dialogue with a user, they could soon be used for canvassing at unprecedented scale[45]. One recent study reported that prompted models could use two-turn interactions to reliably persuade a sample of US citizens on controversial issues such as whether the government should censor materials on the internet or ban fossil fuels to avert climate change[46]. Another showed that after three rounds of interaction, prompted LLMs can reduce participants' beliefs in a range of conspiracy theories (such as the view that 9/11 was a government plot) by up to 20%, and that the effects can endure for two months[47]. These studies hint at the potential power of future LLMs that use persuasive dialogue for political ends, prompting developers to consider mitigations such as increased interpretability and scalable forms of model oversight[48].

**Political polarisation**

In many modern democracies, opinions have become highly polarised, with politics dominated by opposing groups who reject each other's views and values outright. This takes the form of both issue polarisation (highly divergent political perspectives) and affective polarisation (animosity between people with different political affiliations). Polarisation is often blamed on algorithms designed to maximise engagement with digital content that trap users in "filter bubbles" (where their prejudices are constantly reinforced) or "echo chambers" (where they are insulated from the discomfort of contrary views)[49,50]. If LLMs are personalized they may exacerbate this issue, by generating replies that flatter the user's preconceptions, or isolate them from ideologically opposing views. To date, most LLMs are generic (rather than personalised to suit the tastes of individual users) and after fine-tuning mostly provide neutral or diplomatic replies designed to have broad appeal. Nevertheless, there is evidence that even generic LLMs tend to be "sycophantic", or to preferentially express views that may be shared by the user, even if these are untrue. This occurs because human feedback provided during the feedback process tends to reward LLM replies that echo user sentiments[51]. Moreover, some models (such as Replika and Pi) already explicitly tailor outputs based on users' demographics, interests and tastes to make AI more appealing. OpenAI is currently testing a version of the model that remembers user preferences from past conversations[52].

However, it is unclear whether AI personalisation will heighten political polarisation[53]. The view that filter bubbles and echo chambers increases partisanship has come into question[54,55]. Instead, affective polarisation may occur because social media algorithms often encourage the most divisive content to be viewed, attended to and shared[54]. By comparison, publicly available LMs, when fine-tuned to give equanimous perspectives on issues of debate, offer the opportunity to expose users to a spectrum of legitimate opinions, and could nourish public discourse in ways that social media platforms have systematically failed to do[56].

An alternative explanation for heightened partisanship on digital platforms is that voters are allowed to form highly stereotyped perceptions of their political opponents (for example, in the US, Republicans believe that 32% of Democrats identify as LGBTQ, and Democrats believe that 38% of Republicans earn in excess of $250K per year, where in reality the figures are 6% and 2%)[57]. Without careful prompting, LLMs tend to generate caricatured outputs that may exaggerate stereotypical features in exactly this way[58], and thus risk contributing to partisanship by erasing the nuance in the way that people see each other.

**Deliberation and consensus**

In a healthy democracy, people can express diverse opinions, and deliberate in an atmosphere of mutual tolerance and respect. Whilst some fear that AI may be used to weaken or suppress political discussion[59], there is also hope that AI could be used to create healthier spaces for deliberation among citizens. Machine learning tools are already used to moderate content, by identifying insulting, profane or explicit messages online[60] but LLMs may allow us to go a step further, by intercepting uncivil messages and proposing that they are voluntarily withdrawn or rephrased (potentially faster and more reliably than human moderators can). In one study, LLMs were prompted to intervene in political discussions between US voters with opposing views on gun control, by proposing less adversarial message rephrasing. Discussants accepted the proposed wording about two thirds of the time, and when they did so, improvements in perceived conversation quality and democratic reciprocity (the extent to which political opponents report respecting each other's right to hold contrary views) were observed[61]. Another possibility is that LLMs intervention might help amplify voices that are at risk of being marginalised in a discussion. For example, inserting LLMs into mixed gender groups of Afghani citizens discussing contentious political issues was shown to improve the range of ideas contributed by female group members[62]. LLMs also offer new opportunities for improving interactions among citizens in social media or debate platforms, by summarising opinions and optimising the routing of comments between discussants[63], and helping humans themselves become more effective conversation partners[64].

Another promising opportunity is for LLMs to be used to directly assist humans in finding common ground, by facilitating deliberation over issues of legitimate debate. Currently, formal citizens' assemblies allow representative groups of people to gather and debate policy issues or provide collective input into new laws or constitutions[11]. However, organising large-scale in-person events is costly and time-consuming, and face-to-face debate can be susceptible to social desirability biases, where interlocutors are motivated by a desire to win the argument, rather than to reach a mutually acceptable outcome[65]. In one study, LLMs were trained to generate collective statements that maximised endorsement from a group providing private written opinions[66]. These statements were preferred over those written by humans, and helped people converge to a common side of the argument. In another study, LLMs were used in conjunction with a carefully designed process to generate a slate of statements that represent the diversity of opinions in a group, according to a rigorous notion of representation[67]. Building such mathematical guarantees into the outcomes of AI-augmented democratic processes may help alleviate concerns about the biases encoded in LLM summaries, which may otherwise omit essential details or distort the intended meaning[68]. Methods from the field of computational social choice can be used to help LLMs generate more reliable collective outcomes[69].

**Information and misinformation**

LLMs are prone to generate factually unreliable content (or confabulate; this is usually called "hallucination" by AI researchers). Whilst safety fine-tuning pipelines and retrieval techniques are increasingly effective at steering the model towards more accurate statements, model replies can still be poorly sourced, untruthful, or over-confident[70]. Moreover, LLMs are already being deliberately misused to generate misleading content or propaganda. Recent breakthroughs in multimodal generative AI have greatly expanded opportunities for malicious actors to create and manipulate digital content. Some recently deployed models allow users to generate highly realistic audio and video from simple text descriptions, or to alter media in misleading ways. In a political context, this means manipulating multimodal content to portray political rivals in compromising or defamatory ways, producing deceptive campaign videos, and even counterfeiting entire news websites. Already in 2024, deepfake videos have been deployed with obvious intent to shift the electoral calculus in India, Indonesia, Mexico, Pakistan, Slovakia, the US and Taiwan[2]. For example, in Pakistan AI was used to generate a fake video of prime ministerial candidate Imran Khan giving a victory speech from prison, and in Taiwan an AI-generated fake video was released on election day in which a candidate was supposedly endorsed by a former rival – with each of these items receiving hundreds of thousands of views. Whilst there is scant evidence that electoral outcomes were materially affected in these cases, the arrival of hyper-realistic generative content could threaten to rob news media of its "epistemic backstop" - the decisive authority that previously provided by a video or audio recording of a news event. LLMs may also be used as "social bots" on digital platforms, and tasked with spreading false or hyper-partisan content rapidly through networks whilst disguising its AI-based origin[71]. Unfortunately, evidence suggests that. at least in an experimental setting, AI systems may be more convincing when the content they produce is deceptive[34].

AI-generated media is becoming harder to spot. Recent work suggests that AI-generated audio and video[72], images of human faces[73], and tweets[74] may now sometimes be indistinguishable from non-synthetic content. Developers are working on machine learning methods for distinguishing synthetic and human-generated content, and testing "watermarking" techniques that add an invisible signature identifying digital content as generated or altered by AI[75]. However, many current techniques are quite easy to circumvent[76], and researchers acknowledge that watermarking is not a silver bullet for avoiding misinformation[77]. Strong watermarking has been argued to be provably impossible, so these safeguards may only deter relatively unsophisticated threat actors[78,79].

---

[2] https://restofworld.org/2024/elections-ai-tracker/#/pakistan-trump-imran-khan

Over the longer term, repeated exposure to significant volumes of realistic deepfake materials could have a systemic effect on the population's epistemic health. Users are more likely to believe information that is repeated, independent of its plausibility[80], and AI systems offer new and more targeted ways to deluge users with misleading content. Educating people about deepfakes may help, but it can also lead to legitimate content being widely questioned – a phenomenon called the "liar's dividend"[81]. Widespread exposure to deepfakes could thus render people more vulnerable to misinformation campaigns, or breed broader scepticism in online media, as well as traditional news outlets and journalists, across the population[82]. In the worst case, widespread contamination of the digital knowledge commons with fake, deceptive or partisan content could erode public trust in information, eroding our shared understanding of socio-political circumstances or scientific facts [83].

On a more hopeful note, LLMs also provide new opportunities to help users of online platforms to discern truth from falsehood. One example is progress towards automation of fact-checking on digital platforms. This includes chatbot websites themselves: many publicly available LLMs already offer embedded citation, whereby model replies are augmented with hyperlinks that allow users to verify the source of a claim, providing a form of assurance against confabulation[84,85]. LLMs can also be deployed to check the provenance and veracity of claims made on external news or social media sites, a task which is extremely laborious for humans, with some facts taking hours to verify[86]. In one study, ChatGPT was asked to classify more than 20,000 pieces of previously fact-checked news, and was found to agree with human raters about 70% of the time. Interestingly, this rate of agreement was maintained beyond its training cut-off (i.e. to events that it could not possibly have known about) betraying that the model was relying on an estimate of prima facie plausibility to make this judgement rather than actually checking facts as a human might[87]. More reliable fact-checking services may soon be available, but when they arrive, we need to find ways to ensure that they have impact. For example, one study found that participants were just as likely to believe and to share content that ChatGPT had flagged as false as that which it had supposedly verified[88], and people generally mistrust ChatGPT as a source of political information[89].

## Material Impacts

Democracy is an idealised principle of governance, but in modern societies its material realisation relies partly on technology, including digital technology[4,90]. As well as influencing how citizens consume and digest political information, technology shapes how individuals and groups can participate in collective decision-making in a democracy, by debating, protesting, lobbying, polling, funding, or voting. It determines how elected representatives communicate policies and principles with citizens, and how policy is implemented by the bureaucratic machinery of state. AI is the transformative technology of the 21st century, and so it naturally has an impact on the *materiality* of democracy – the infrastructure that supports the democratic process in society.

In 2024, a year in which so many countries go to the polls, there has been an uptick of concern that AI could be deployed to disrupt elections. There is the worry that malicious actors, including rival nation-states, could use AI-generated misinformation to distort political campaigning. This includes the use of AI for "hack and leak" operations, whereby private accounts (e.g. email) may be compromised to obtain information that can be disseminated to portray political rivals in an unflattering light, or to otherwise unfairly tip the political balance. For example, the Russian spy agency Star Blizzard is thought to have targeted western politicians and journalists over a sustained period, as well as directly attacking the UK Electoral Commission[91]. A common strategy is to use spear-phishing campaigns, in which individuals are maliciously targeted with highly personalised messages, typically over email. LLMs may be able to assist with this process, by crafting bespoke deceptive messages based on personal details (e.g. scraped from social media profiles). For example, one study showed how ChatGPT can be used to generate spear phishing messages tailored to British Parliamentarians[92].

**Election-related misuse**

The material practice of democracy relies on elections and referenda being free and fair – eligible voters should be able to cast their vote unhindered. As voters turn to AI with questions about elections, developers need to ensure that LLMs provide accurate, up-to-date information about eligibility and voter registration, polling station access, voter ID, and other election rules. At present, this is not always the case. For example, one study tested the accuracy of leading proprietary and open source models on practical queries about electoral participation in the US, finding that over half of replies were inaccurate[93]. The problem may be particularly acute for those models (like the free-to-use version of ChatGPT) which do not use real-time internet queries to obtain up-to-date information for replies, and thus risk providing outdated advice (although deployed models are increasingly fine-tuned to encourage users to seek information from authoritative sources).

Unfortunately, attempts have been made to misuse AI to influence voter turnout.. In one well-publicised example, generative AI was used to synthesise an automated telephone message (or robocall) which appeared to feature President Joe Biden discouraging voters from participating in the 2024 New Hampshire primary. Given the ease with which such deepfake materials can be generated – using a short snippet of genuine audio and a few dollars – efforts to use generative AI to sow confusion among voters and officials could grow. In the near future, heightened personalisation of messages to individuals could exacerbate this risk, for example with robocalls that feature tailored disinformation about eligibility to vote (e.g. based on past felony convictions). Tracking and disabling tools that allow these malicious activities is becoming increasingly difficult.

Another vulnerability is voter registration, which is already a battleground issue in many US states. According to recent reports[94] a tool called EagleAI, which purports to identify fraudulent voter activity, is being deployed by activists to query or reject legitimate registrations (especially from minorities in contested wards) on the basis of unreliable evidence. EagleAI has been approved for voter roll maintenance in at least one Georgia county, potentially giving it the power to arbitrate over thousands of registration challenges[95]. A related risk is that AI's ability to generate content at scale is used to deliberately overwhelm electoral infrastructure, undermining the credibility of the democratic process or suppressing voter participation *en masse*. In the US, many states have seen a huge surge in voter records requests (sometimes running to millions of documents) made under freedom of information laws, in an apparent attempt to disrupt legitimate election audit processes[96]. AI can be used to accelerate this sort of disruptive activity. For example, EagleAI also allows partisan groups to file mass voter challenges (attempts to strip large numbers of registrants of their vote) on the basis of limited evidence. By automating this process, activists using AI can lodge an overwhelming number of challenges immediately before the review deadline, rendering officials powerless to overturn them.

In a well-functioning democracy, citizens are free to engage politically as journalists, activists, and politicians, free from the threat of online violence and abuse. However, hyper-partisan groups often spread disinformation with a view to discouraging some constituencies from political participation When this happens, evidence indicates that women and minorities are disproportionately targeted [97]. For example, female Democrats are ten times more likely to receive online abuse than their male counterparts[98]. Generative AI is used to craft and disseminate increasing volumes of gendered disinformation and defamation[99], and to synthesise deepfake pornographic images, sometimes targeted at political leaders or activists[100].

**Augmenting political decision making**

Democratic representatives are empowered to make choices on behalf of citizens, but to do so they need to access and process relevant information. Currently, politicians rely heavily on experts to brief them on relevant issues (such as the Congressional Research Service in the US, and other stakeholders and advocates). It has been proposed that LLMs might support human political decision-making, helping politicians summarise vast bodies of data, brainstorming policy initiatives, or writing draft legislation[101]. This could allow legislatures to write, debate, and pass more effective bills, or aid in the insertion of "micro-legislation," minor and subtle text that changes the effect of laws[102], as well as aiding in the detection of loopholes. AI may even start to draft entire pieces of legislation (even if based on human desiderata) – in November 2023 the legislature of Porto Alegre, Brazil, passed the first law written entirely by an LLM[103]. If AI can help politicians respond better to citizen's needs, this could bolster their perceived legitimacy as democratic representatives.

AI systems can also potentially enhance conduits of communication between legislators, public servants, and the electorate. For example, LLMs can produce well-structured texts or oratory which could help politicians communicate ideas more clearly to their constituents. In turn, AI may open new avenues for people to feed back their views to government. LLMs are already being used for more effective election polling, harnessing social media data to make microscale predictions about voting intentions that match or exceed those from statistical models used by professional pundits[104,105]. AI systems may also facilitate civic education, by helping voters inform themselves about the issues that most concern them, and which parties best represent their interests. LLMs may also be used to empower citizens by providing easier routes to learn about their rights, or to help them navigate state bureaucracy and legal processes. For example, the UK has launched a conversational agent that responds to queries about topics covered on government webpages, including the details of forthcoming elections[3].

Much more is possible. AI systems have already been shown to provide balanced summaries of the opinions expressed by small groups of people[66,67], but in newer LLMs with longer context lengths (the number of input tokens on which they can condition their output) this automated opinion digest could potentially be scaled to groups of thousands or more, providing a new, LLM-based mechanism for governments to ascertain what citizens think and want. The summarisation process could even be conditioned on demographics, allowing insight into how both majority and minority groups may respond to a political decision[106]. In theory, if participants' beliefs are modelled accurately, it could be possible to hold an "election" for every decision, in which AI agents vote on behalf of stakeholders. This approach has been successfully piloted in the context of sensitive decisions involving food allocation, which must balance considerations of fairness and efficiency[107], although whether AI systems should serve as human proxies in high-stakes settings remains a controversial topic[108]. Whilst we should be wary of naïve techno-solutionism[109], advanced AI systems invite us to reimagine how democracies might work for the better.

---

[3] https://www.govgpt.uk/

## Foundational Impacts

Democracy is based on a set of shared values and principles that undergird democratic institutions and offset the burden of democratic participation[110]. As a number of political analysts have noted, if these norms are eroded, democracies may 'backslide', or gravitate toward authoritarianism[111]. Aside from the epistemic and material impacts discussed above, AI could either corrode or bolster the foundations of democracy – to either accelerate, or guard against, democratic backsliding.

One major concern is that AI will serve to concentrate excessive power in the hands of political leaders or parties. According to one view, democracy flourished in the 20th century because the technological landscape favoured decentralised economies and polities, where power is distributed across diverse groups and individuals[112]. However, AI could be used to dramatically streamline 21st century state bureaucracies towards centralized governance, and potentially strengthening authoritarian forms of governance. Commentators have speculated that AI intrinsically favours 'turn-key authoritarianism'[113] or even 'tyranny'[114].

Faith in democratic institutions can be undermined by a perception that the political process is rigged to create winners and losers. For example, many elected governments are perceived as being unresponsive to the demands of the majority, and catering selectively to the few. Populist parties with anti-democratic agendas are poised to exploit these grievances for their own political advantage. There are fears that AI will accelerate this trend by increasing inequality, especially in developed nations[115]. For example, new capabilities exhibited by LLMs could lead to the displacement of some sectors within the labour market, including administrative and creative industry jobs that were previously thought likely to be spared automation[116]. However, the precise impact that AI will have on the economy and composition of the workforce remains uncertain, with some forecasting new opportunities for job creation[117,118]. For example, emerging studies show that equipping workers with LLMs tends to bring the skill levels of lower skilled workers in line with their more highly trained counterparts[119,120], which could imply that AI will help create a more level playing field in the workforce. A related concern is that without appropriate governance, the wealth generated by this technological revolution could become concentrated in the hands of a few multinational corporations, in a handful of countries, who are building AI and distributing its services[121]. Moreover, governments may struggle to keep up with the pace and complexity of private technology development and so critical choices about the way AI shapes societies may be made by corporations instead of democratic polities[122]. If so, this could reinforce the perception that democracy's cherished liberal principles have evolved to serve elites rather than society as a whole.

In a democracy, representatives and public officials need to be accountable for their actions. If politicians fail to deliver, they can be voted out of office. AI could undermine this principle by blurring lines of accountability when policies fail – because it is unclear whether human or machine made the final decision[123]. In consumer settings, we have already seen an airline attempt to designate an LLM as "a legal entity with responsibility for its own decisions" – in order to pass the blame to AI for erroneous customer advice. As AI systems become embedded in the machinery of government, this creates new opportunities for blame to be deflected and accountability obscured. Increased automation of the levers of state could also have other unwanted secondary effects, such as disempowering citizens with inflexible bureaucracy, subjecting everyone to a relentless world of "computer says no"[125]. We should be wary that advances in AI do not presage a descent into "algocracy" – government by algorithm, in which humans are wholly or partly removed from the loop[126].

However, if administered properly AI holds the potential not only to conserve but also to enhance the foundational elements of democracy. AI might well bolster confidence in democratic government by creating wealth, enhancing productivity, and improving health, education and digital infrastructure. Cultural commitments to democracy could also be strengthened by the epistemic and material impacts described earlier, including improvements in government service delivery, facilitation of communication between constituents and elected officials, and scaling of public democratic deliberation. In the limit, AI may provide an opportunity to update the social contract underlying representative democracy, by expanding the role of citizen input in policymaking, up to and including the design of democratic institutions and processes[66,107,127]. Such reinventions of our democratic institutions may be critical if our existing systems are unprepared for handling the dramatic changes that unfold as advanced AI becomes ever more ubiquitous in our society[128].

## Conclusions and outlook

AI is the most significant technology of our times. However, as its impact expands, it is becoming clear that it will increasingly interface with another of humanity's most important inventions: democratic self-governance. In this brief review, we have surveyed the possible implications of this encounter for the future of democracy.

AI will present specific challenges to democracy at multiple levels: epistemic, material and foundational. However, AI also holds out potential affirmative opportunities. Our analysis suggests that neither exuberant optimism nor despairing pessimism is an appropriate stance. Instead, what is called for is clear-eyed and persistent efforts to shape both the design of AI technology and the design of democratic institutions so that they fit together well, yielding democratic benefits from AI while preventing democratic harms. If we plan carefully, we should be able to assure—and even enhance—our democratic future.

*Disclaimer: Any opinions presented in this paper represent the personal views of the authors and do not necessarily reflect the official policies or positions of their organisations.*